\documentclass[10pt,twocolumn,superscriptaddress,floats,showpacs,nobalancelastpage,longbibliography,prb]{revtex4-2}

\usepackage{amsmath,amsfonts, amssymb, amsthm, dsfont}
\usepackage{yfonts}
\usepackage{bm}
\usepackage{mathrsfs}
\usepackage{graphicx}
\usepackage{verbatim}
\usepackage[pdfencoding=auto, psdextra]{hyperref}
\hypersetup{linktocpage}
\hypersetup{colorlinks=true,citecolor=blue,linkcolor=blue, urlcolor=blue}
\usepackage{enumitem}
\usepackage{wasysym}
\usepackage{xcolor}
\usepackage{tikz}
\usepackage{bbold}
\usepackage[caption=false]{subfig}
\usepackage{times}

\renewcommand{\vec}[1]{{\mathbf #1}}
\newcommand{\ket}[1]{|#1\rangle}

\newcommand{\comments}[1]{}

\def\U{\mathrm{U}(1)}

\def\Z{\mathbb{Z}}

\makeatletter
\def\l@subsubsection#1#2{}
\makeatother

\definecolor{OliveGreen}{cmyk}{0.64, 0, 0.95, 0.40}

\definecolor{tyler}{rgb}{1,.3,0}

\begin{document}
\title{Chiral spin liquid in a $\Z_3$ Kitaev model}

\author{Li-Mei Chen}
\affiliation{Guangdong Provincial Key Laboratory of Magnetoelectric Physics and Devices, School of Physics, Sun Yat-sen University, Guangzhou 510275, China}
\affiliation{State Key Laboratory of Optoelectronic Materials and Technologies, Sun Yat-sen University, Guangzhou 510275, China}

\author{Tyler D. Ellison}
\affiliation{Department of Physics, Yale University, New Haven, CT 06520-8120, USA}

\author{Meng Cheng}
\email{m.cheng@yale.edu}
\affiliation{Department of Physics, Yale University, New Haven, CT 06520-8120, USA}

\author{Peng Ye}
\email{yepeng5@mail.sysu.edu.cn}
\affiliation{Guangdong Provincial Key Laboratory of Magnetoelectric Physics and Devices, School of Physics, Sun Yat-sen University, Guangzhou 510275, China}
\affiliation{State Key Laboratory of Optoelectronic Materials and Technologies, Sun Yat-sen University, Guangzhou 510275, China}

\author{Ji-Yao Chen}
\email{chenjiy3@mail.sysu.edu.cn}
\affiliation{Guangdong Provincial Key Laboratory of Magnetoelectric Physics and Devices, School of Physics, Sun Yat-sen University, Guangzhou 510275, China}
\affiliation{Center for Neutron Science and Technology, Sun Yat-sen University, Guangzhou 510275, China}

\date{\today}

\begin{abstract}

We study a $\mathbb{Z}_3$ Kitaev model on the honeycomb lattice with nearest neighbor interactions. Based on matrix product state simulations and symmetry considerations, we find evidence that, with ferromagnetic isotropic couplings, the model realizes a chiral spin liquid, 
characterized by a possible $\mathrm{U}(1)_{12}$ chiral topological order. This is supported by simulations on both cylinder and strip geometries. On infinitely long cylinders with various widths, scaling analysis of entanglement entropy and maximal correlation length suggests that the model has a gapped 2D bulk. The topological entanglement entropy is extracted and found to be in agreement with the $\mathrm{U}(1)_{12}$ topological order. On infinitely long strips with moderate widths, we find the model is critical with a central charge consistent with the chiral edge theory of the $\mathrm{U}(1)_{12}$ topological phase. We conclude by discussing several open questions.

\end{abstract}
\maketitle

\section{Introduction} 
Quantum spin liquids are highly-entangled phases of matter, characterized by fractionalized excitations and emergent gauge fields~\cite{Misguich2005, Savary2016, Zhou2017}. The physics of spin liquids is exemplified by the Kitaev honeycomb model, which is a remarkable example of an exactly-solvable model that exhibits both gapped and gapless $\text{spin-liquid}$ phases~\cite{Kitaev2006}. Furthermore, when the interactions are isotropic and the time-reversal symmetry is broken (either by a small magnetic field or a three spin interaction), the model is in a chiral spin liquid (CSL) phase, characterized by the Ising topological order. The CSL phase supports non-Abelian Ising anyons in the bulk, which can be utilized for topological quantum information processing~\cite{Nayak2008}. Via the bulk-boundary correspondence, it also has gapless edge states described by a chiral Ising conformal field theory (CFT)~\cite{Kitaev2006}, which can provide the smoking-gun evidence for experimental realizations of the Kitaev model through thermal Hall measurements~\cite{Kasahara2018}.

Besides rapid progress towards realizing the Kitaev model experimentally~\cite{Trebst2022, Takagi2019, Micheli2006, Slagle2022, Verresen2022}, generalizations of this model, including relevant non-Kitaev type interactions~\cite{Liu2018, Wang2019}, higher spins~\cite{Dong2020, Jin2022} and various lattices~\cite{Yao2007, Chulliparambil2020} have also been under intensive investigation. These generalizations share the common property that the fractionalized degrees of freedom contain fermions. One notable exception is the generalized Kitaev model on spin-$1$ degrees of freedom proposed in Ref.~\cite{Barkeshli2015}, which we shall call the $\Z_3$ Kitaev model hereafter. In the $\Z_3$ Kitaev model, the local spin degrees of freedom fractionalize into $\Z_3$ parafermions. Using a coupled-wire construction, it was suggested in Ref.~\cite{Barkeshli2015} that the $\Z_3$ Kitaev model could support a $\Z_3$ parafermion topological order. 

However, in contrast to the orginal Kitaev model, which can be solved exactly using a Majorana fermion representation~\cite{Kitaev2006, Feng2007}, the $\Z_3$ Kitaev model is not exactly solvable. This is due to the fact that the Hamiltonian terms are quadratic in the $\Z_3$ parafermion operators, implying that the system is interacting as opposed being a free theory. Although it has been shown that a $\Z_3$ toric code topological phase exists in the highly anisotripic parameter region of the $\Z_3$ Kitaev model, much less is known about the case of isotropic couplings, which could provide a fertile ground for exotic spin liquids.

Remarkably, despite the $\Z_3$ Kitaev model not being exactly solvable, it possesses certain generalized symmetries that enforce the existence of a set of anyonic excitations, under the assumption of a spectral gap. These symmetries are sufficient for predicting only a subset of the anyons present in the theory. Thus, more systematic work is needed to clarify the nature of the phase at the isotropic point. 


In this work, we study the $\Z_3$ Kitaev model numerically using matrix product states to complement the symmetry considerations. Based on results on cylinder and strip geometries, we found evidences that the model at the ferromagnetic isotropic point is gapped and realizes a chiral spin liquid phase. Our numerically measured topological entanglement entropy and chiral central charge suggest an exotic $\mathrm{U}(1)_{12}$ chiral topological order in this phase, in contrast to the $\Z_3$ parafermion topological order suggested in Ref.~\cite{Barkeshli2015}.

\section{$\Z_3$ Kitaev Model and symmetries} 
\begin{figure}[htbp]
\centering
    \includegraphics[width=0.96\columnwidth]{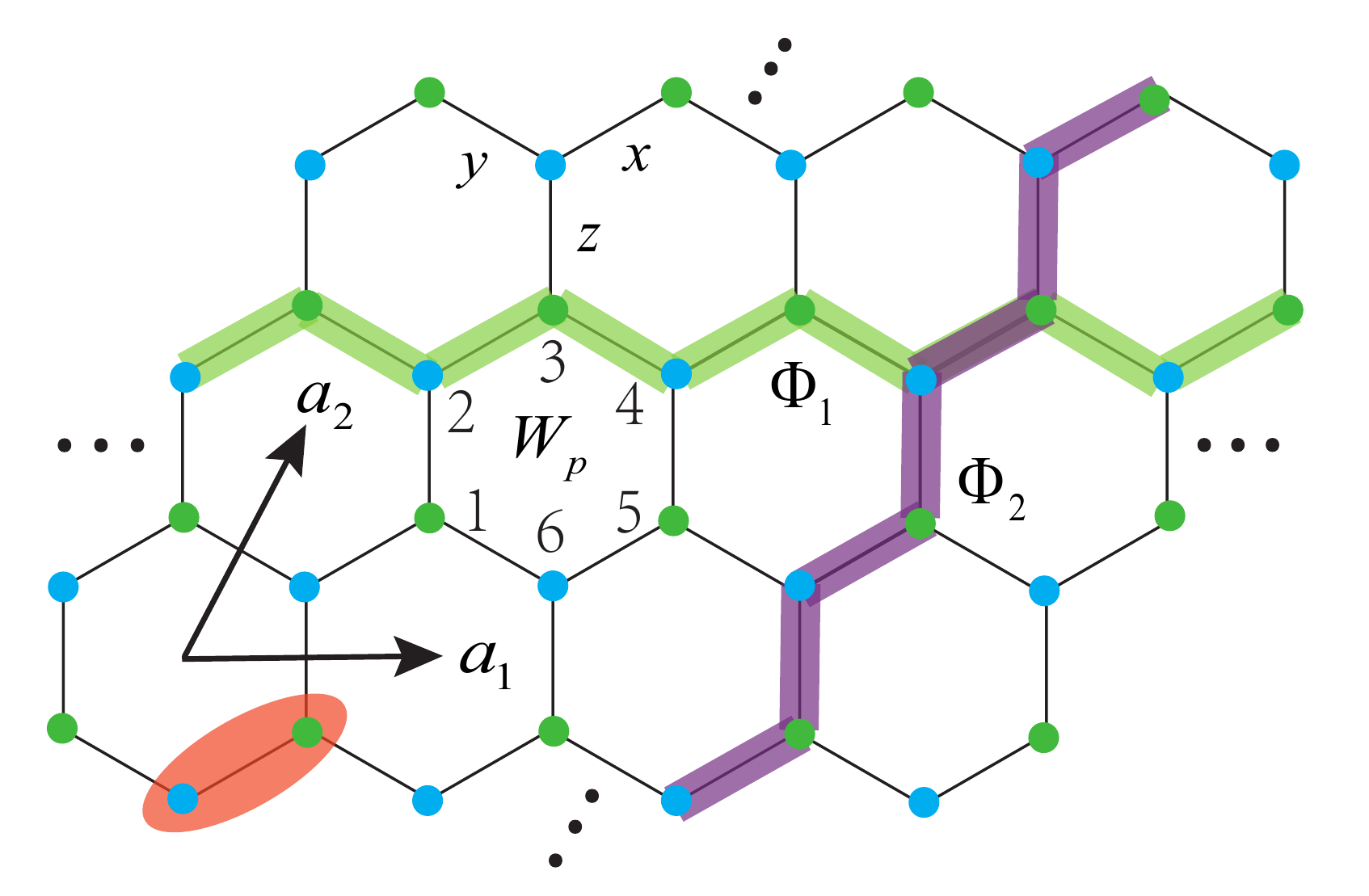}
    \caption{Honeycomb lattice for the $\Z_3$ Kitaev model, generated by translations along $\mathbf{a}_1=(1,0)^\mathrm{T}$ and $\mathbf{a}_2=(\frac{1}{2},\frac{\sqrt{3}}{2})^\mathrm{T}$ direction with a two-site unit cell. The $A$ ($B$) sub-lattice is shown by green (blue) dots. Three different classes of links are labeled by $x$, $y$, $z$. The length of the system in the $\mathbf{a}_1$ ($\mathbf{a}_2$) direction is denoted as $L_x$ ($L_y$). Conserved quantities $W_p$, $\Phi_1$, $\Phi_2$ are indicated.}
\label{fig:lattice}
\end{figure}

The $\Z_3$ Kitaev model of Ref.~\cite{Barkeshli2015} is defined on a hexagonal lattice with a three-dimensional Hilbert space at each vertex. We label the links of the lattice by $\alpha=x,y,z$, according to Fig.~\ref{fig:lattice}. The Hamiltonian is given by
\begin{equation} 
\label{eq: Z3 Hamiltonian}
    H=-\sum_{\alpha=x,y,z}J_\alpha \sum_{\langle ij \rangle \in \alpha\text{-links}}T_i^\alpha T_j^\alpha+\text{h.c.}.
\end{equation}
Here, $T_i^x, T_i^y$ and $T_i^z$ are unitary operators supported at the vertex $i$, defined by the following relations: $(T_i^x)^3=(T_i^y)^3=(T_i^z)^3=1$, $T_i^z\equiv (T_i^x T_i^y)^{\dag}$, and $T_i^x T_i^y=\omega T_i^y T_i^x$, where $\omega \equiv e^{\frac{ 2 \pi i}{3}}$. Explicit expressions of $T^{x,y,z}$ are given in Appendix~\ref{sec:explicit_form}. Note that, the Hamiltonian in Eq.~\eqref{eq: Z3 Hamiltonian} is complex instead of being real, suggesting the time reversal symmetry is broken. This is in stark contrast to the original Kitaev model.

Similar to the original Kitaev model, the Hamiltonian in Eq.~\eqref{eq: Z3 Hamiltonian} has an extensive number of conserved quantities. For each hexagon $p$, the conserved quantity $W_p$ is defined as $W_p = (\omega T_1^x T_2^yT_3^zT_4^xT_5^yT_6^z)^\dag$, where the operators are labeled by the vertices of $p$ as shown in Fig.~\ref{fig:lattice}. It can be checked that these $W_p$'s are mutually commuting. More generally, for any contractible path $\gamma$, one can define a closed string operator along $\gamma$ by multiplying the $W_p$'s over the plaquettes in the region enclosed by $\gamma$. 

The $\Z_3$ Kitaev model also has conserved quantities supported along non-contractible paths. For example, the string operator $\Phi_1$, supported along a path $\gamma_1$ in the $\mathbf{a}_1$-direction (Fig. \ref{fig:lattice}), is defined as $\Phi_1=\prod_{i\in \gamma_1} (T^z_i)^{s_i}$, where $s_i=1 (-1)$  for $i\in A (B)$ sublattice. Similarly, the string operator $\Phi_2$ is given by $\Phi_2=\prod_{i\in \gamma_2}(T^y_i)^{s_i}$, where $\gamma_2$ is the path along the $\mathbf{a}_2$-direction. These conserved quantities satisfy the relation $\Phi_1\Phi_2=\omega\Phi_2\Phi_1$, implying that every energy level of the Hamiltonian has a three-fold degeneracy. We note that the set of conserved quantities generated by $W_p$, $\Phi_1$, $\Phi_2$ form a generalized $\Z_3$ symmetry, known as a $\Z_3$ 1-form symmetry~\footnote{Here, the symmetry is distinct from the 1-form symmetry introduced in Ref.~\cite{Gaiotto2015}, since the symmetry operators are not truly topological~\cite{Qi2021}.}. See Refs.~\cite{Gaiotto2015, Qi2021, Ellison2022} for more details.

The conserved quantities described above in fact imply that every gapped phase captured by the Hamiltonian in Eq.~\eqref{eq: Z3 Hamiltonian} must host anyons of the so-called $\Z_3^{(2)}$ anyon theory~\cite{Ellison2022, Bonderson2012}. 
To make this explicit, we consider truncations of the conserved string operators. If the string operator is truncated  to an open path $\gamma$, then the truncated string operator $W(\gamma)$ fails to commute with $H$ only at the two end points of $\gamma$. This suggests that, if the system is in a gapped phase, the open string operator $W(\gamma)$ creates anyonic excitations at its endpoints (see Ref.~\cite{Ellison2022} for a more precise statement). Motivated by this, we can assign an Abelian anyon theory to the algebra of string operators. The properties of the associated anyons (i.e., fusion rules, exchange statistics, and braiding) are entirely determined by the algebra of string operators.

In particular, since the closed string operators satisfy the relation $W(\gamma)^3=1$, the anyons obey $\Z_3$ fusion rules. The exchange statistics $\theta$ can be computed from segments of string operators using the methods of Refs.~\cite{Levin2003, Kawagoe2020}. We find that the anyons created by the open string operator $W(\gamma)$ have exchange statistics $\theta=e^{\frac{4\pi i}{3}}$. The $\Z_3$ fusion rules and $\theta=e^{\frac{4\pi i}{3}}$ exchange statistics uniquely determine that the Abelian anyon theory is $\Z_3^{(2)}$. The total quantum dimension of the $\Z_3^{(2)}$ anyon theory is $\mathcal{D} = \sqrt{3}$ and the chiral central charge is $c_- = -2 \text{ mod }8$.

In general, the anyon theories of the gapped phases of the $\Z_3$ Kitaev model factorize as $\mathcal{C}\boxtimes \Z_3^{(2)}$, where $\boxtimes$ denotes that anyon theories $\mathcal{C}$ and $\Z_3^{(2)}$ are independent~\cite{Muger2003}.
The total quantum dimension of the product anyon theory $\mathcal{C}\boxtimes \Z_3^{(2)}$ factorizes as $\mathcal{D} = \mathcal{D}_\mathcal{C}\sqrt{3}$, where $\mathcal{D}_\mathcal{C}$ is the total quantum dimension of $\mathcal{C}$. The chiral central charges are additive, so the total chiral central charge is $c_- = c_-' - 2 \text{ mod 8}$, with $c_-'$ being the chiral central charge of $\mathcal{C}$.

As a concrete example, in the anisotropic limit $J_z \gg J_{x},J_y$, the model can be mapped to a $\Z_3$ toric code~\cite{Barkeshli2015}, which corresponds to $\mathcal{C}=\Z_3^{(1)}$. In this case, $\mathcal{D}_\mathcal{C} = \sqrt{3}$ and $c'_- = 2 \text{ mod }8$. Therefore, the total quantum dimension of $\Z_3^{(1)}\boxtimes \Z_3^{(2)}$ is $\mathcal{D}=3$ and the total chiral central charge is $c_-=0 \text{ mod }8$.

The Hamiltonian also has a global $\mathbb{Z}_3$ symmetry, generated by $U = \prod_{i\in A} T_i^z \prod_{j\in B} (T_j^z)^{\dag}$, with $A$, $B$ denoting the two sublattices. The symmetry  
$U$ can be naturally understood in terms of the $\Z_3$ 1-form symmetry, as it is the product of $\Phi_1$ operators over all of the paths along the $\mathbf{a}_1$ direction.

Before presenting the numerical results, we comment on the interplay between the conserved quantities and the global symmetries of the system on a cylindrical geometry. In particular, we consider a cylinder that is periodic in the $\mathbf{a}_2$-direction and has a circumference of $L_y \neq 0 \text{ mod }3$. We also assume, in agreement with the numerical results below, that the ground states are eigenstates of the $W_p$'s with an eigenvalue of $W_p = \omega^n$, where $n \neq 0 \text{ mod }3$. 
From this, we draw the following two conclusions: \textbf{(i)} the global $\Z_3$ symmetry is spontaneously broken, so if the symmetry is enforced on the ground state, then it will be a Schr\"odinger's cat state, \textbf{(ii)} if the symmetry is not enforced and we consider a ground state that breaks the symmetry, then the translation symmetry in the $\mathbf{a}_1$-direction must also be broken.

To see \textbf{(i)}, we first note that, since $L_y \neq 0 \text{ mod }3$, the string operator $\Phi_2$, which wraps around the cylinder, is charged under the global $\Z_3$ symmetry. To show that the $\Z_3$ symmetry is spontaneously broken, it is sufficient to show that $\Phi_2$ has long-range correlations. Letting $\Phi_{2,x}$ be the string operator $\Phi_2$ supported on sites sharing the same coordinate of the $\mathbf{a}_1$-axis (denoted as $x$), we compute the correlator $\langle \Phi_{2,x}\Phi_{2,x'}^\dag \rangle$. This is non-vanishing in the limit of a large separation between $x$ and $x'$, because $\Phi_{2,x}\Phi_{2,x'}^\dag$ is equivalent to a product of $W_p$ operators over the region between $x$ and $x'$. Thus, using that the ground states are eigenstates of the $W_p$'s, we have $\langle \Phi_{2,x}\Phi_{2,x'}^\dag \rangle = \omega^{n|x-x'|L_y}$. A consequence is that, if the global $\Z_3$ symmetry is enforced, the system is in a Schr\"odinger's cat state.

To see \textbf{(ii)}, we notice that $\langle \Phi_{2,x} \rangle = \langle \Phi_{2,x}\Phi_{2,x+1}^\dagger \Phi_{2,x+1} \rangle$. Since $\Phi_{2,x}\Phi_{2,x+1}^\dagger$ is equivalent to a product of $W_p$'s in the column between $x$ and $x+1$, we have $\langle \Phi_{2,x}\rangle=\omega^{n L_y}\langle\Phi_{2,x+1}\rangle$. This implies that, if the global $\Z_3$ symmetry is not enforced and $\langle \Phi_{2,x} \rangle \neq 0$, then the translation symmetry in the $\mathbf{a}_1$-direction is spontaneously broken. 


\section{Results on cylinder geometry}
\begin{figure}[htb]
\centering
    \subfloat{\includegraphics[width=0.492\columnwidth]{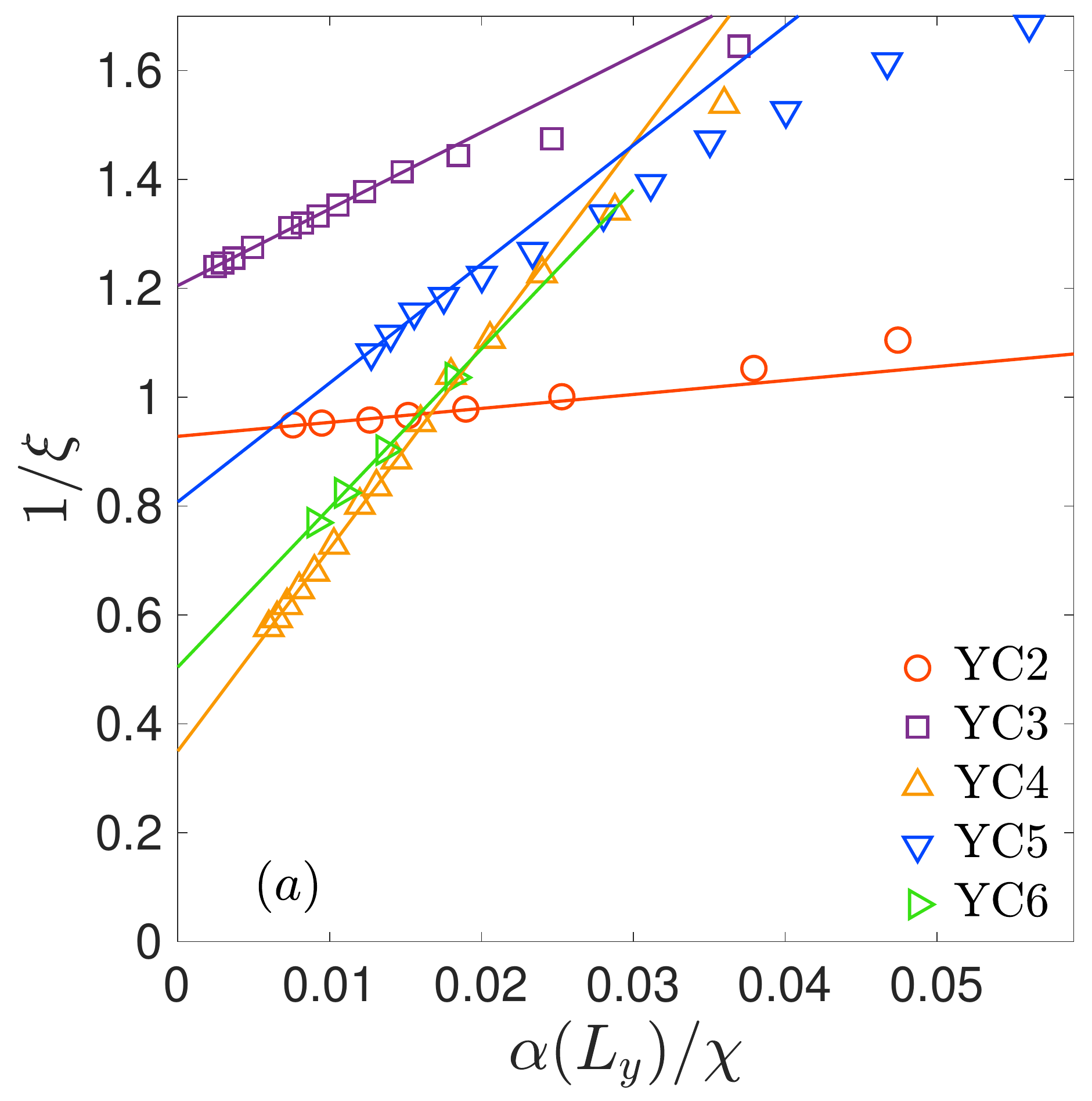}}
    \subfloat{\includegraphics[width=0.48\columnwidth]{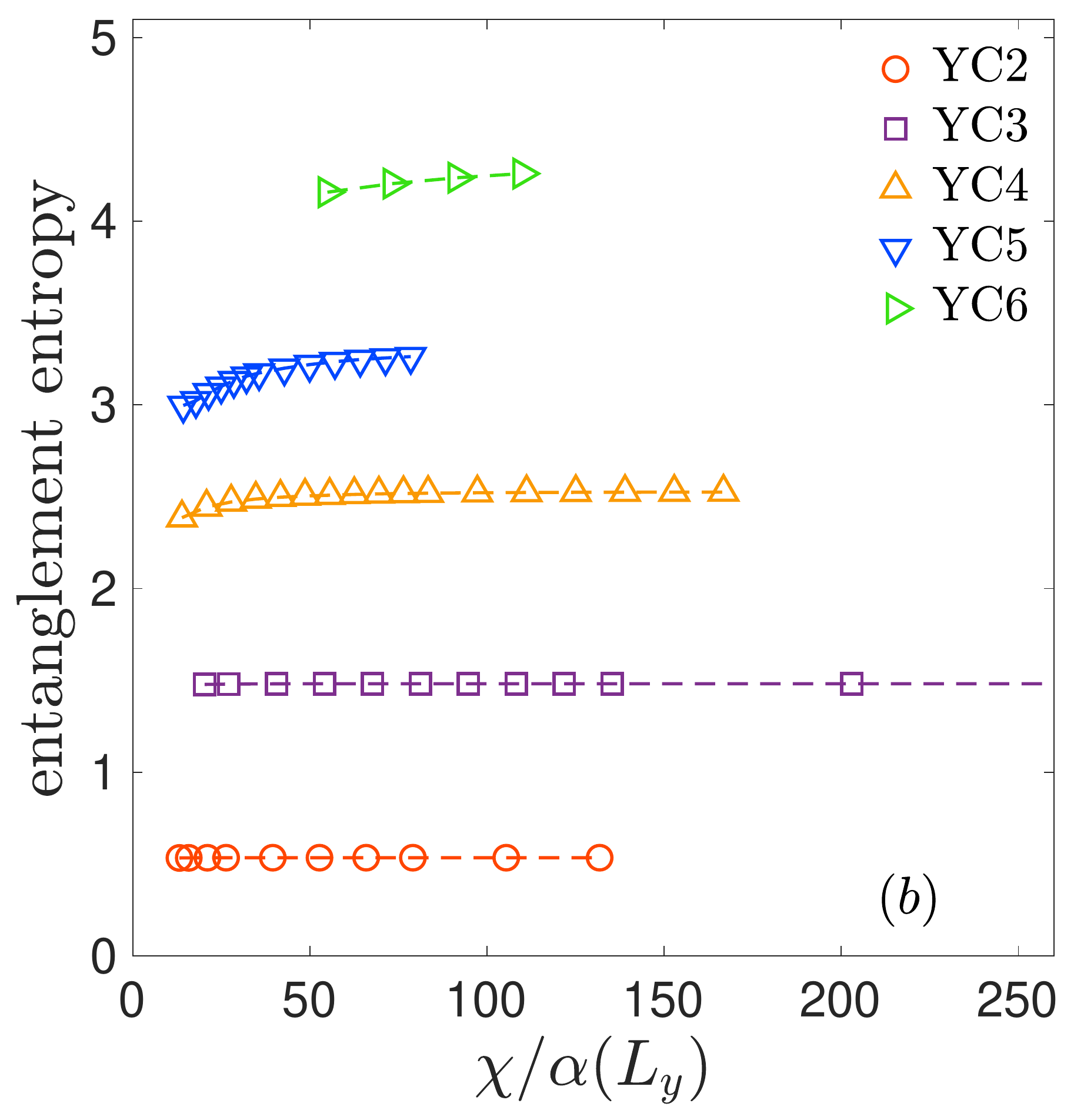}}
\caption{Correlation length and half cylinder entanglement entropy for infinitely long cylinders with YC$L_y$ geometries and $L_y$ between $2$ and $6$. 
To compare results for all computed cylinders, we have rescaled the bond dimension $\chi$ by $\alpha(L_y)=\mathrm{exp}(L_y/1.5)$. In (a), solid lines are linear fit using the last few points with large $\chi$. We note that the $y$-intercept is nonzero, implying that for large bond dimensions, the correlation length saturates to a finite value. In (b), dashed lines are guide to the eyes.
}
\label{fig:cylinder}
\end{figure}

Let us start with the cylinder geometry, i.e. periodic boundary condition along the $\mathbf{a}_2$-direction and varying circumferences $L_y$. Following Ref.~\cite{Yan2011}, we denote this class of clusters as YC$L_y$. Hereafter, we focus on the parameters $J_x=J_y=J_z=1$, and use the matrix-product-state (MPS) based infinite density matrix renormalization group (iDMRG) algorithm to probe bulk properties~\cite{McCulloch2008, Schollwock2011, White1992, Stoudenmire2012}. The main physical observables to consider include the maximal correlation length $\xi$ (extracted from transfer matrix of the MPS) and entanglement entropy $S=-\mathrm{tr}\left(\rho_A \ln \rho_A \right)$, where $\rho_A$ is the reduced density matrix for the left half of infinitely long cylinders. The results are shown in Fig.~\ref{fig:cylinder}.

For a gapped system, we expect the maximal correlation length $\xi$ and entanglement entropy $S$ for width-$L_y$ cylinder saturate to a finite value in the infinite bond dimension limit. This is clearly observed for YC$2$ and YC$3$ cylinders, as shown in Fig.~\ref{fig:cylinder}. In these two cases, the truncation errors with the largest bond dimension considered ($\chi_\mathrm{max}=500, 3000$ for YC$2$, YC$3$, respectively) are less than $10^{-10}$, in agreement with the rapid saturation behavior. For YC$4,5,6$ geometry, since the required MPS bond dimension $\chi$ for a given accuracy grows exponentially with $L_y$, it becomes challenging to achieve full saturation of $\xi$. Nevertheless, the entanglement entropy clearly saturates with increasing $\chi$, and the truncation error is around $3\times 10^{-7}$ ($3\times 10^{-5}$, $6\times 10^{-5}$) with $\chi_\mathrm{max}=2400 (2200, 6000)$ for YC$4 (5,6)$, respectively. A linear in $1/\chi$ fit indicates that the true correlation length in the infinite $\chi$ limit is less than $3$ lattice spacing, which indeed agrees with existence of a spectral gap in these geometries.

Some remarks are in order. Firstly, for all the data shown in Fig.~\ref{fig:cylinder}, the ground states are simultaneous eigenstates of $W_p$ and $\Phi_2$. And on all considered cylinders, the ground state has a uniform $\Z_3$ flux with $\langle W_p \rangle = \omega^2$ on each plaquette, in contrast with the flux-free ground state in the original Kitaev model~\cite{Kitaev2006}. This was further confirmed on a small torus using exact diagonalization and on finite cylinders ($L_y=3, L_x=6, 12$ and $L_y=4, L_x=12$) using density matrix renormalization group (DMRG). As mentioned above, the nontrival flux means that the unit cell of the MPS in iDMRG simulation is at least three columns for $L_y\neq 0$ mod $3$. For YC$3$ and YC$6$ geometries, we have checked that MPS ansatzs with unit cell of one column and three columns give the same result.

Secondly, as discussed above, when $L_y$ is not a multiple of 3, if the $\Z_3$ symmetry is imposed on the MPS, the ground state found by iDMRG is a Schr\"odinger cat state with diverging correlation length. Thus we do not impose the $\Z_3$ symmetry for YC$2,4,5$ cylinders. For YC$3, 6$, we have checked that the results obtained using $\Z_3$ symmetry with total charge $Q=0$ are identical to those without using this good quantum number, for relatively small bond dimensions~\footnote{We have checked via ED, that the ground state on a 12-site torus has $\Z_3$ charge 0}.
This allows us to use the $\Z_3$ symmetry for YC$3,6$ to reach larger bond dimension. The $\Z_3$ symmetry operator in Eq.~\eqref{eq: Z3 Hamiltonian} is not uniform on every site and one can apply a charge conjugation transformation on one sublattice to make it easy to exploit for MPS calculations, see Appendix~\ref{sec:explicit_form} for further details about the transformation. 

While it is difficult to make definite conclusions, our results are consistent with a gapped state in the 2D bulk. If this is indeed the case, then the bulk must be topologically ordered, due to the $\Z_3$ 1-form symmetry. In the following we study possible signatures of the topological order.

A defining feature of intrinsic topological order is long-range entanglement in the ground state~\cite{Levin2006, Kitaev2006TEE}. Given any region $A$ in a gapped state, the entanglement entropy generally takes the form $S_A=\mathrm{const.}\times |\partial A|-S_\text{top}+\cdots$, where $|\partial A|$ denotes the length of the boundary of $A$, $S_\text{top}$ is a universal constant known as the topological entanglement entropy (TEE), and $\cdots$ includes terms vanishing for large $A$. For planar system, $S_\text{top}=\ln \mathcal{D}$, where $\mathcal{D}$ is total quantum dimension of the underlying topological order. Choosing the entanglement cut along the cylinder circumference~\cite{Jiang2012}, $S_\text{top}$ depends on which ground state the system is in. If the ground state has a definite anyon flux $a$ through the cylinder (equivalently, as a quasi-1D system the state is short-range correlated), then $S_\text{top}=\ln \frac{\mathcal{D}}{d_a}$, with $d_a$ the quantum dimension of anyon type $a$. (Note that for Abelian anyon theories, such as $\Z_3^{(2)}$, we can take $d_a = 1$ for all $a$.)

\begin{figure}[htb]
\centering
    \includegraphics[width=0.6\columnwidth]{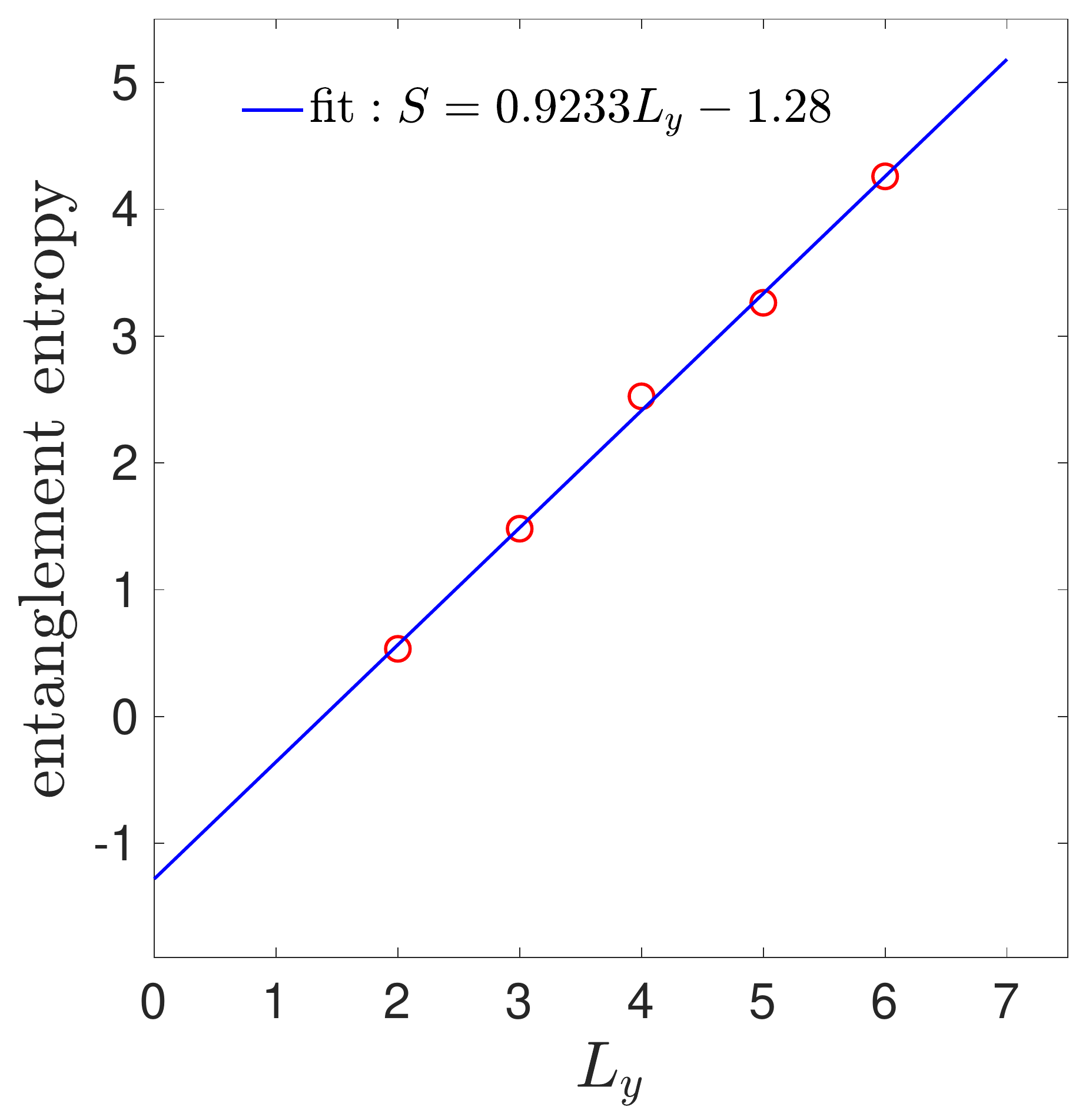}
\caption{Entanglement entropy versus $L_y$ on infinitely long cylinders. For each $L_y$, we take the value with largest $\chi$.
A linear fit reveals a finite TEE $S_\text{top}=1.28\pm 0.34$.
}
\label{fig:TEE}
\end{figure}

In Fig.~\ref{fig:TEE}, we show the entanglement entropy versus cylinder width $L_y$ (for $L_y$ between $2$ and $6$), from which a nonzero TEE $S_\text{top}=1.28\pm 0.34$ is extracted. Assuming that the ground states are in the identity sector of the topological order~\footnote{This can be justified by various random initial states giving the same result}, we have $S_\text{top}=\ln \mathcal{D}$, which gives $\mathcal{D}\approx 3.6$. 

Besides entanglement entropy, chiral topological order could also be revealed by entanglement spectrum (ES) of bipartition of the cylinder into two halves~\cite{Li2008}. Indeed, the Li-Haldane conjecture states that the low energy content of the ES is described by the CFT governing the physical edge properties, which has also been extensively used in characterizing chiral topological phase~\cite{Cincio2013,Chen2021}. Here, the conserved quantities $W_p$'s lead to extensive degeneracy in the ES~\cite{Yao2010, Shinjo2015}, making it hard to identify further structures numerically (such as chiral CFT tower). Some examples of ES are shown in Appendix~\ref{sec:ES}. Nevertheless, we can probe edge properties by studying the system on a strip geometry.

\section{Results on strip geometry} 
The above study on a cylindrical geometry reveals bulk properties of the $\Z_3$ Kitaev model, such as the existence of a spectral gap and the total quantum dimension.
Using the bulk-edge correspondence, the edge theory of this model would be given by a 1+1D chiral CFT with central charge equal to the chiral central charge of the topological order. This can be revealed by putting the system on a strip geometry, where the chiral and anti-chiral gapless modes at the two edges are weakly coupled by tunneling through the bulk, with the coupling strength being exponentially suppressed by the bulk gap. Thus for a sufficiently wide strip, one would expect to see a 1+1D CFT (at least within a relatively large length scale), which could be tested via finite entanglement scaling.

Besides the criticality, we show in Appendix~\ref{sec: degeneracy} that there is an additional $3^{L_x}$-fold degeneracy on the strip geometry, due to non-commuting string-like conserved quantities that extend between the bottom and top of the strip. In the strip geometry, there are also conserved quantities $W_{p, \text{boundary}}$ corresponding to the plaquettes that would connect the top and bottom boundaries if the system were made periodic in the $\mathbf{a}_2$-direction. Numerically, we find that the expectation value of $W_{p, \text{boundary}}$ is $\omega$ or $\omega^2$ with the same energy (exact or in the infinite $\chi$ limit), for all the strip widths we have considered $L_y=2,3,4,5$.

In our numerical simulations in  the strip geometry, it is also important to ensure that the obtained ground states are eigenstates of $\Phi_2$, to avoid long-range correlations from spontaneous symmetry breaking. In practice, we find that this can be achieved by starting the simulation with a small penalty term $-\Phi_{2,x}\Phi_{2,x+1}\Phi_{2,x+2} + \mathrm{h.c.}$ (which commutes with the Hamiltonian Eq.~\eqref{eq: Z3 Hamiltonian}), and the associated ground state serves as the initial state for further optimization. Alternatively, we can add $-\Phi_{2,x}$ with additional phase factor $1,\omega,\omega^2$ and its hermitian conjugate to the Hamiltonian, to enforce a specific $\langle \Phi_2 \rangle$ configuration. In addition, we can also pin the conserved quantity $W_{p, \text{boundary}}$ to be $\omega$, by adding a suitable pinning term.

\begin{figure}[hbtp]
\centering
    \subfloat{\includegraphics[width=0.47\columnwidth]{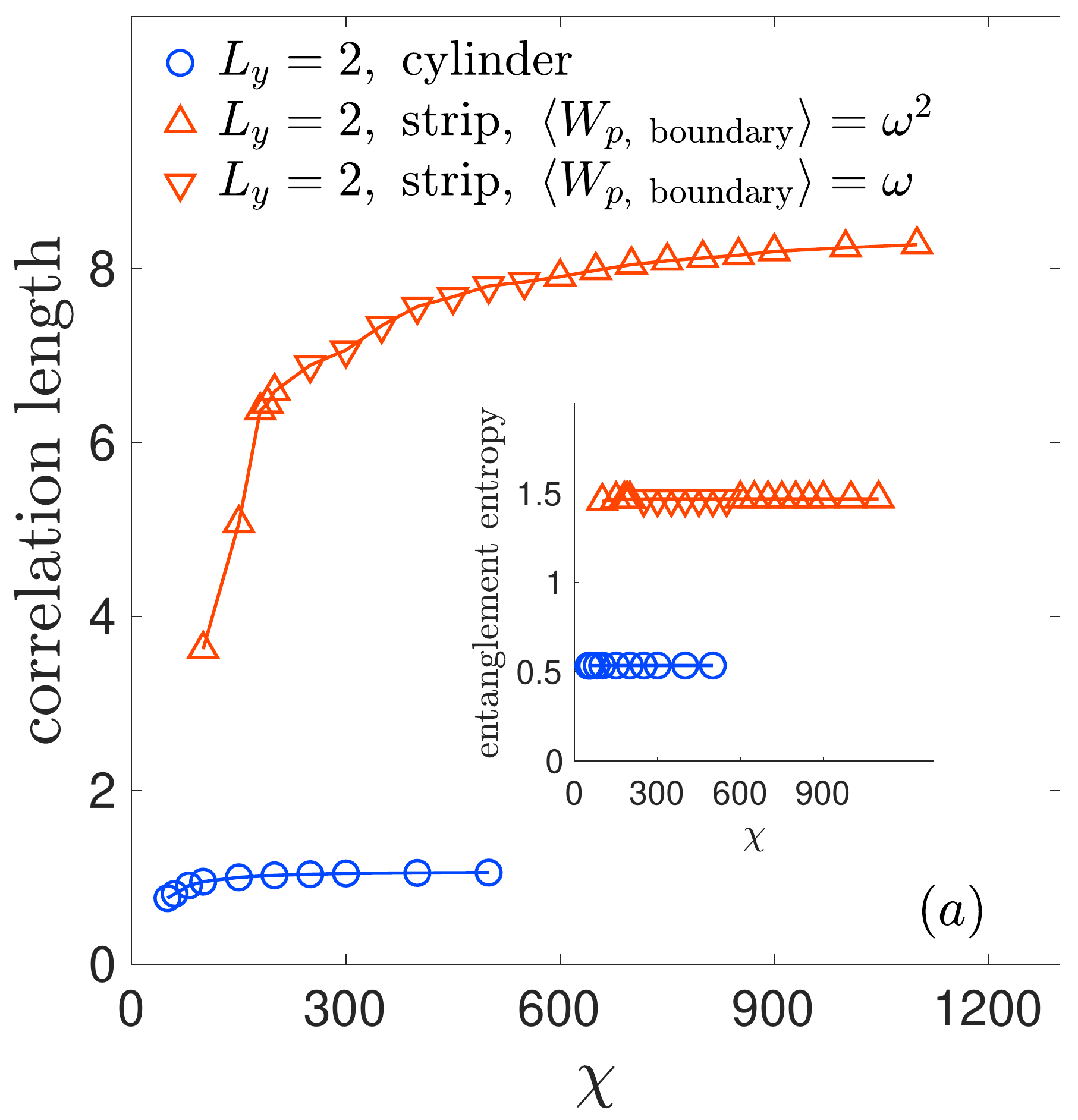}}
    \subfloat{\includegraphics[width=0.495\columnwidth]{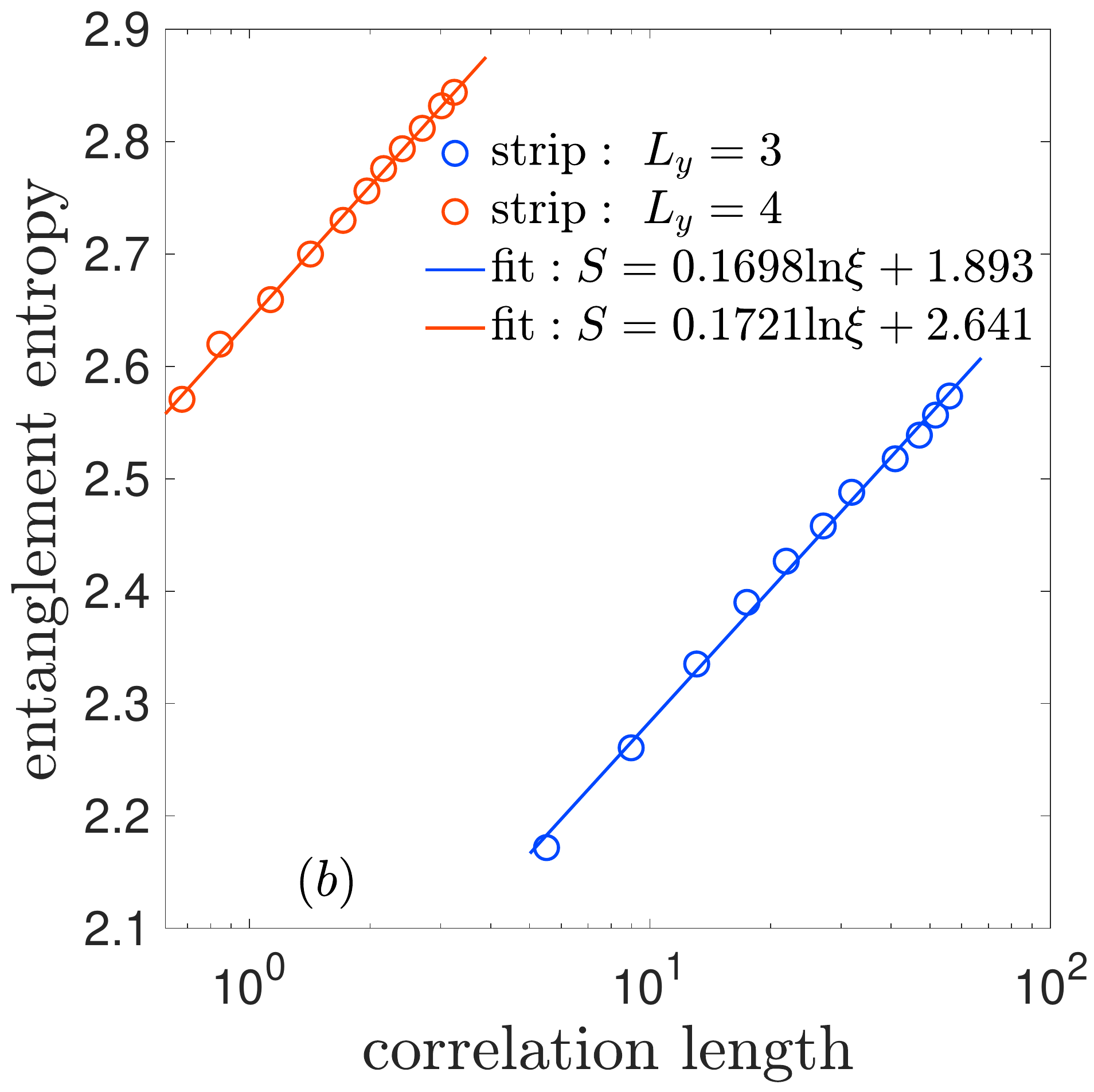}}
\caption{(a) For the $L_y=2$ strip, both the maximal correlation length (main panel) and entanglement entropy (inset) are converging as functions of bond dimension $\chi$, with values significantly larger than those on the YC$2$ cylinder. (b) For strip with width $L_y=3$, $4$, a diverging behavior of correlation length and entanglement entropy versus $\chi$ was found, which we fit using the finite entanglement scaling.
}
\label{fig:strip_Jz_-1_Ly_2_3_4_Lx_3}
\end{figure}

\begin{figure}
    \centering
    \subfloat{\includegraphics[width=0.48\columnwidth]{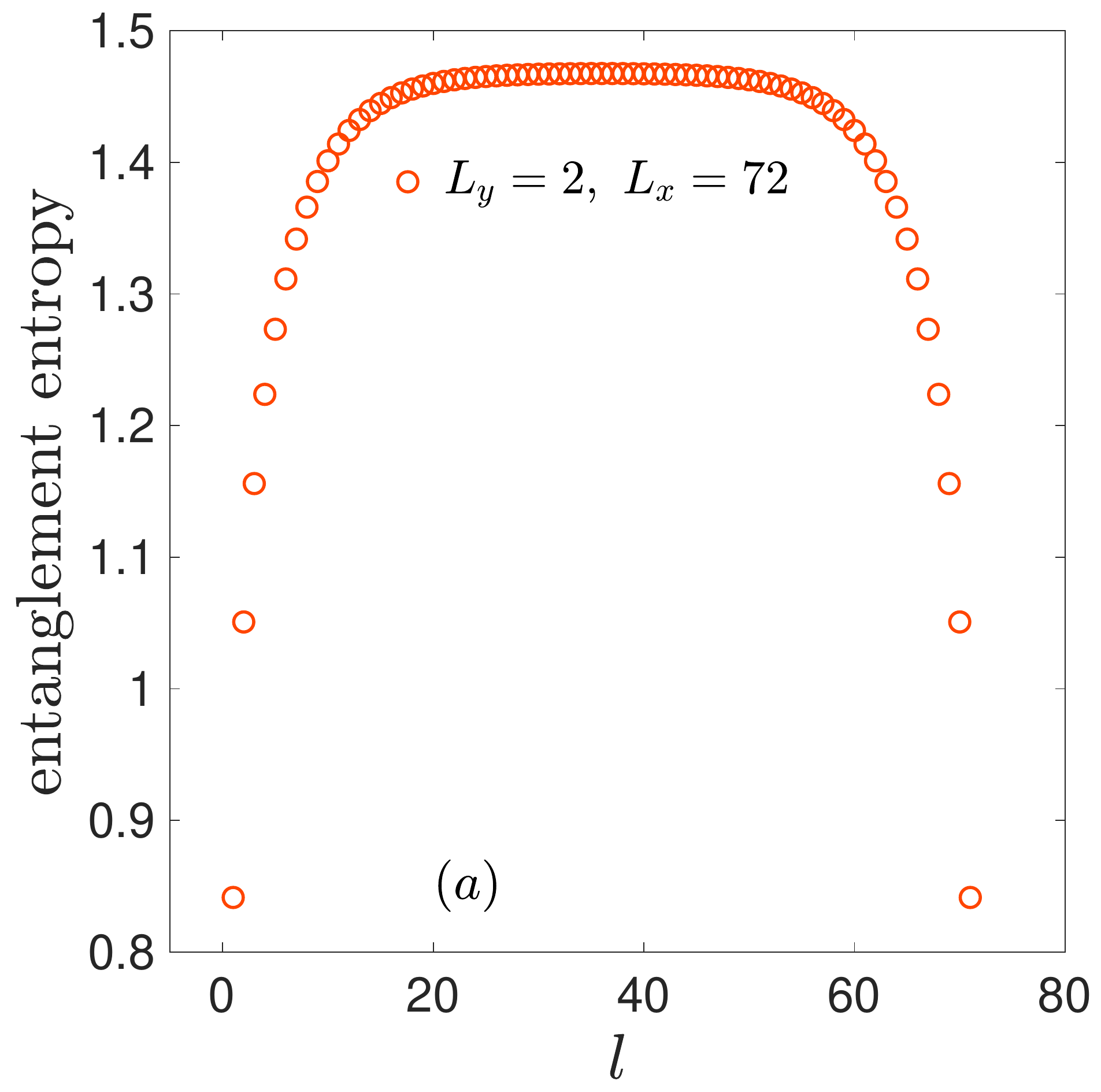}}
    \subfloat{\includegraphics[width=0.485\columnwidth]{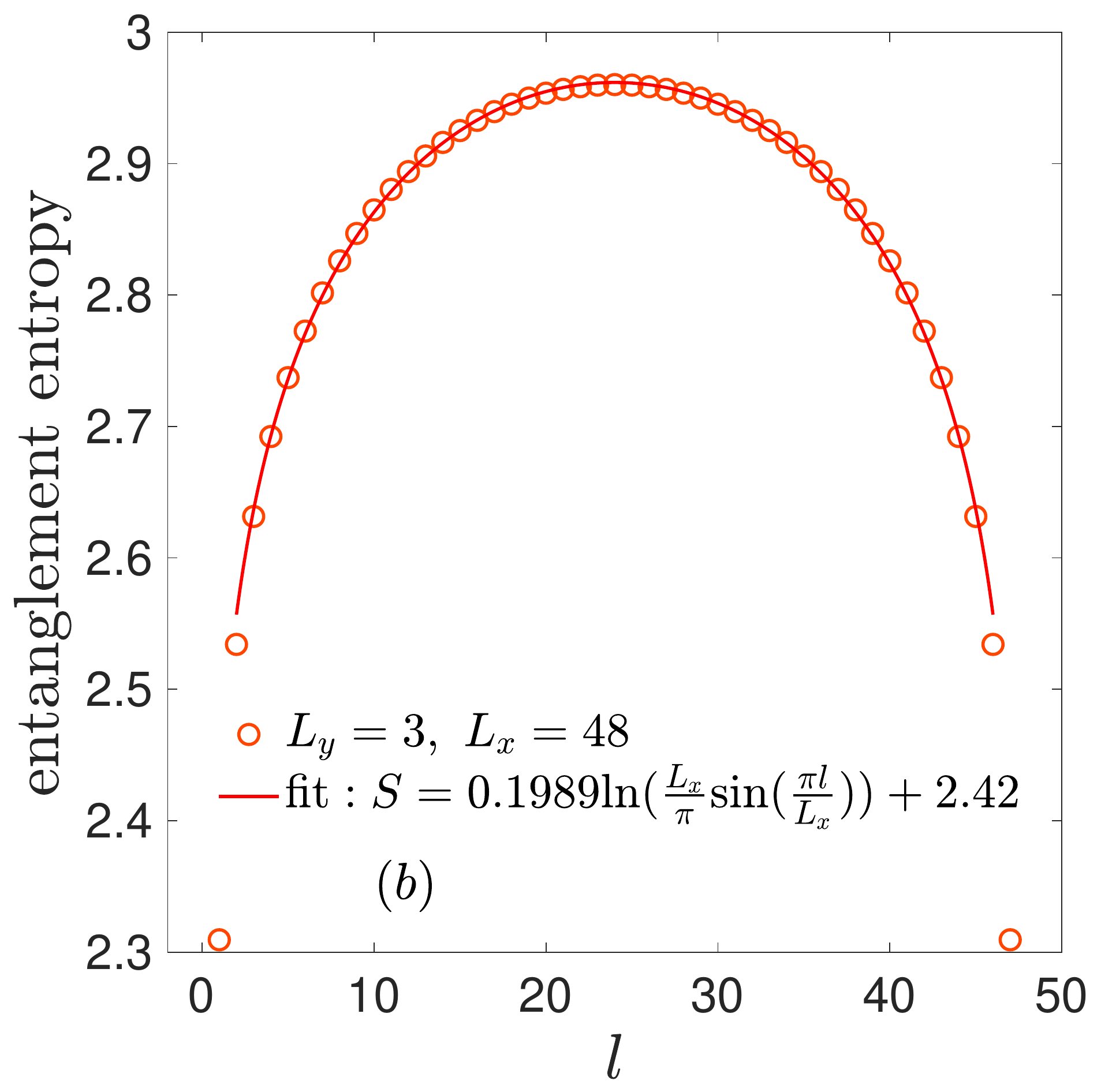}}
    \caption{Entanglement entropy of a length $l$ subsystem on the left of a finite sized strip with total length $L_x$ and width $L_y$. (a) $L_y=2, L_x=72$. The bond dimension is $\chi=1000$, and truncation error is $5.4\times 10^{-11}$. With increasing $l$, the entanglement entropy clearly saturates. (b) For $L_y=3, L_x=48$, the bond dimension is $\chi=2000$, and truncation error is $4.4\times 10^{-7}$. Using Calabrese-Cardy formula, a central charge $c\approx 1.2$ is extracted.}
    \label{fig:finite_size_strip}
\end{figure}

It turns out that, for the strip of width $L_y=2$, both correlation length and the half-strip entanglement entropy saturate with increasing bond dimension $\chi$, as shown in Fig.~\ref{fig:strip_Jz_-1_Ly_2_3_4_Lx_3}(a), suggesting a small finite gap. The relatively large value of the correlation length, however, indicates that the system may potentially become more critical with increasing $L_y$. For $L_y={3, 4}$, we observe that the system indeed becomes gapless (within the correlation length set by the bond dimension), as evidenced from the increasing correlation length and half-strip entanglement entropy with $\chi$ showing no sign of saturation. From a fit of the half-strip entanglement entropy and correlation length~\cite{Pollmann2009} (shown in Fig.~\ref{fig:strip_Jz_-1_Ly_2_3_4_Lx_3}(b)), we extract a central charge $c=1.019 (1.033)$ for $L_y=3 (4)$. We thus conclude that $L_y=3$ and $4$ strips have $c=1$, which we interpret as the chiral central charge of the bulk topological order. 

In addition, as a consistency check, we have also performed DMRG studies on finite-size strips, where both the width and length (denoted as $L_y$ and $L_x$, respectively) are finte. In this case, to fix the spectrum degeneracy, we have also added $\Phi_2$ (and its Hermitian conjugate) into the Hamiltonian, while taking care of the relation between $\Phi_2$ and $W_p$.
Typical results (among results with various bond dimension and strip length) are shown in Fig.~\ref{fig:finite_size_strip}.
It turns out, on a sufficiently long strip with width $L_y=2$, the entanglement entropy clearly saturates with increasing subsystem length (denoted as $l$), as shown in Fig.~\ref{fig:finite_size_strip}(a). On the contrary, on a width $L_y=3$ strip with length $L_x=48$, a finite central charge $c\approx 1.2$ can be extracted using the Calabrese-Cardy formula (see Fig.~\ref{fig:finite_size_strip}(b))~\cite{Calabrese2004}, which is slightly larger than the central charge on infinitely long strip possibly due to finite size effect.

Further increasing $L_y$, it becomes numerically expensive to measure the central charge. Nevertheless, on $L_y=5$ we observe that the the entanglement entropy and maximal correlation length do not show any sign of saturation with increasing $\chi$ (data shown in Appendix~\ref{sec: Ly_5_strip}), consistent with the strip being gapless. Since the models are fully gapped on the cylinders, 
 the most natural explanation of the critical properties is that they all come from the gapless edge states described by the same $c=1$ CFT, as one would expect in a strip of chiral topological state.

\section{Discussion and conclusion}

In summary, we have studied the ferromagnetic isotropic point of the $\Z_3$ Kitaev model proposed by Barkeshli \textit{et al.}~\cite{Barkeshli2015}. Through a detailed study on both cylinder and strip geometries with various widths, we have identified signatures of a chiral spin liquid. 

We now discuss possible topological order based on the numerical results. From the cylinder and strip calculations we have observed that
\begin{enumerate}
    \item All cylinders up to $L_y=6$ have finite correlation lengths, which show no sign of diverging as $L_y$ increases.
    \item The $L_y=3$ and $4$ strips appear to be critical with $c\approx 1$.
    \item From the entanglement entropy results on cylinders we find the TEE is $S_{\rm top}=1.28\pm 0.34$.
\end{enumerate}

The simplest explanation of these results is that the ground state is gapped in the bulk, and has gapless edge modes. Suppose the 2D topological phase has a CFT edge theory with both left- and right-moving modes, and the corresponding central charges are $c_L$ and $c_R$. The chiral central charge of the 2D phase is given by $c_-=c_L-c_R$, and, assuming the top and bottom edges are decoupled the strip has central charge $c=c_L+c_R$ . Notice that here $c_L$ are $c_R$ are both non-negative. If both are positive, for a unitary CFT it is known that $c_L, c_R$ is at least $1/2$ (the Ising CFT). So we have found essentially three options: 
\begin{equation}
    (c_L, c_R)=(1,0), (0,1), (1/2,1/2).
\end{equation}
The last case means $c_-=0$ and the edge is just a (non-chiral) Ising CFT. While this scenario can not be excluded, we find it unlikely since there is no obvious mechanism to prevent the edge modes from opening a gap. Thus, we focus on the other two cases, where the edge modes are fully chiral.

According to classification of chiral $c=1$ CFTs~\cite{Ginsparg:1987eb}, these theories come in two types: 1) $\U_{2n}$ theories with $n$ a non-zero integer, often referred to as the ``circle branch". 2) Orbifolds of the circle branch, including $\Z_2$ orbifolds of $\U_{2n}$ and three exceptional cases (orbifolding non-Abelian groups in SU(2)$_1$). We will focus on the circle branch, which will turn out to be more relevant for this system. The $\U_{2n}$ CFT can be described by a chiral compact boson. They can arise as the edge theory of a 2+1D chiral topological phase, the bulk of which is described by the $\U_{2n}$ anyon theory. An example is the $\nu=\frac12$ bosonic Laughlin state (also known as the Kalmeyer-Laughlin state) with $\U_2$ chiral edge CFT. Let us enumerate the anyon content of the $\U_{2n}$ theory. The different anyon types can be labeled by an integer defined mod $2n$, denoted as $[a]$ where $a\in \Z$, with $[0]$ representing the trivial excitations.  $[a]$ has exchange statistics $e^{\frac{i\pi a^2}{2n}}$. The fusion rule of anyons is given by addition, so $[a]\times [b]=[a+b]$. The anyon types form a $\Z_{2n}$ group under fusion, with $[1]$ as the generator. Using the notations in \cite{Bonderson2012}, the theory is denoted as $\Z_{2n}^{(1/2)}$.

However, we have also established through the exact 1-form symmetry of the lattice model that the bulk anyon theory must contain the $\Z_3^{(2)}$ as a subtheory, which places strong constraint on the value of $n$. First, to have $\Z_3^{(2)}$ as a subtheory, $\Z_3$ must be a subgroup of $\Z_{2n}$, meaning that $n$ must be a multiple of $3$, so we write $n=3m$. The unique $\Z_3$ subgroup in $\Z_{6m}$ is $\{[0], [2m], [4m]\}$. The self statistics of the generator $[2m]$ is $e^{\frac{i\pi}{6m}(2m)^2}=e^{\frac{2\pi i}{3}m}$, therefore to match $\Z_3^{(2)}$, $m$ must be 2 mod 3, and $n$ must take the form $n=3(3k+2)$ for $k\in \Z$. We note that a similar analysis can exclude the orbifold branch.

To further constrain the possible values of $k$, we consider the TEE. It is known that the TEE for the $\U_{6(3k+2)}$ theory is $\ln\sqrt{6\big|3k+2\big|}$. Compared with the fitted TEE $S_{\rm top}=1.28\pm 0.34$, we find that $k=0$ (the $\U_{12}$ theory) has the closet TEE $\ln \sqrt{12}\approx 1.24$. We note that $k=-2$ (the $\U_{-24}$ theory with TEE $\ln \sqrt{24}\approx 1.58$) is also compatible, due to the relatively large errorbar on the numerically computed TEE. While $\U_{-24}$ can not be excluded based on our data so far, we believe the fact that $\U_{-24}$ has many more anyons and the TEE is close to the upper bound set by the errorbar makes it less likely to be the the actual ground state topological order.  Based on this analysis, we identify $\U_{12}$ as the most likely candidate topological order.

Let us discuss the properties of the $\U_{12}=\Z_3^{(2)}\boxtimes \Z_4^{(3/2)}$ anyon theory. The anyon content has been discussed above. It is easy to check that $[4]$ generates $\Z_3^{(2)}$, and $[3]$ generates the $\Z_4^{(3/2)}$ part. This theory has 12 degenerate ground states on torus. Starting from one of them, the others can be obtained by applying the corresponding Wilson loop operators. Since the $\Z_3^{(2)}$ Wilson loops commute with the Hamiltonian, the ground states are at least three-fold degenerate (as well as all the excited states). The degeneracy associated with $\Z_4^{(3/2)}$ is however not expected to be exact and is generally split by quasiparticle tunneling. Indeed, we have managed to find a topologically quasi-degenerate ground state on a width-3 cylinder (discussed in Appendix~\ref{sec: topo_deg}). Nevertheless, due to the high topological degeneracy, finding all topologically degenerate ground states seems to be a daunting task.



Before closing, let us also mention the open questions regarding this model.
First, we should caution the readers that the conclusion of a $\mathrm{U}(1)_{12}$ chiral spin liquid relies on the value of TEE, which still shows a large errorbar (see caption of Fig.~\ref{fig:TEE}).
Second, it is worthwhile to better understand microscopic mechanism that can stabilize the $\mathrm{U}(1)_{12}$ topological phase in this type of models,  for example using a coupled wire analysis.  Last, currently, only the central charge is obtained from a strip geometry. It would be of great interest to extract the low energy excitations on the same geometry, which in principle would allow a more complete characterization of the CFT. Given the rapid advancement of tensor network methodology~\cite{Zou2018,Vanderstraeten2019}, this is indeed a promising avenue to further pin down the nature of this chiral topological phase.

\section{Acknowledgements} 
We acknowledge conversations with Johannes Hauschild, Xiao-Yu Dong, Hui-Ke Jin and Hong-Hao Tu. The MPS calculations were performed using the TeNPy Library (version 0.9.0)~\cite{Hauschild2018}. L.M.C. and P.Y. were supported by NSFC Grant  No.~12074438, Guangdong Basic and Applied Basic Research Foundation under Grant No.~2020B1515120100, and the Open Project of Guangdong Provincial Key Laboratory of Magnetoelectric Physics and Devices under Grant No.~2022B1212010008. M.C. acknowledges support from NSF under award number DMR-1846109. J.-Y.C. was supported by Open Research Fund Program of the State Key Laboratory of Low-Dimensional Quantum Physics (Project No.~KF202207), Fundamental Research Funds for the Central Universities, Sun Yat-sen University (Project No.~23qnpy60), a startup fund from Sun Yat-sen University, the Innovation Program for Quantum Science and Technology 2021ZD0302100, Guangzhou Basic and Applied Basic Research Foundation (grant No.~2024A04J4264),
and National Natural Science Foundation of China (NSFC) (Grant No.~12304186).
This work was also supported by the Fundamental Research Funds for Central Universities (22qntd3005). 
The Calculations reported were performed on resources provided by the Guangdong Provincial Key Laboratory of Magnetoelectric Physics and Devices, No.~2022B1212010008.

\appendix

\section{Local operators and charge conjugation}
\label{sec:explicit_form}

For completeness, here we first show the explicit expressions of the $T^{x,y,z}$ operators.
\begin{equation}
T^x = \left( \begin{smallmatrix} 0 & 0 & 1\\
1 & 0 & 0\\
0 & 1 & 0\\
\end{smallmatrix} \right),
T^y = \left( \begin{smallmatrix} 0 & \omega^2 & 0\\
0 & 0 & \omega\\
1 & 0 & 0\\
\end{smallmatrix} \right),
T^z = \left( \begin{smallmatrix} 1 & 0 & 0\\
0 & \omega & 0\\
0 & 0 & \omega^2\\
\end{smallmatrix} \right),
\end{equation}
where $\omega=\mathrm{exp}(i 2 \pi/3)$.

As mentioned in the main manuscript, the model in the main text has a global $\Z_3$ symmetry, generated by
\begin{equation}\nonumber
 U = \prod_{i\in A} T_i^z \prod_{j\in B} (T_j^z)^{\dag}.
\end{equation}
For the purpose of using symmetries to lower the computational cost, it is useful to perform a charge conjugation transformation on one of the sublattice, say $B$ sublattice, to transform the generator $U$ into a uniform form. This transformation is given by 
\begin{align}
C  T^x C^\dag = (T^x)^{\dag},\\ \nonumber
C  T^y C^\dag = \omega^2(T^y)^{\dag}, \\
C  T^z C^\dag = (T^z)^{\dag}, \nonumber
\end{align}
where $C = \left(\begin{smallmatrix}1&0&0\\0&0&1\\ 0 & 1 & 0\end{smallmatrix}\right)$.

Under this transformation, the Hamiltonian is transformed into
\begin{equation}\nonumber
\begin{split}
H=&-\sum_{\langle i j\rangle \in x\text{-links}} J_x T_{i}^{x} (T_{j}^{x})^{\dag} -\sum_{\langle i j\rangle \in y\text{-links}} J_y \omega^2 T_{i}^{y} (T_{j}^{y})^{\dag} \\
&-\sum_{\langle i j\rangle \in z\text{-links}} J_z T_{i}^{z} (T_{j}^{z})^{\dag}+ \text{h.c.},
\end{split}
\end{equation}
and the plaquette operator $W_p$ becomes
\begin{equation}\nonumber
\begin{split}
W_p=(T_1^x)^{\dag} T_2^y (T_3^z)^{\dag} T_4^x (T_5^y)^{\dag} T_6^z.
\end{split}
\end{equation}
The string operators $\Phi_1$ and $\Phi_2$ are transformed accordingly.

As noted in the main manuscript, the $\Z_3$ symmetry generator commutes with the string operator $\Phi_2$ only
when $L_y$ is a multiple of 3. Thus one can use this $\Z_3$ symmetry in MPS-based calculations only for $\mathrm{mod}(L_y, 3)=0$. For other widths with $\Z_3$ symmetry expolited, a Schr\"odinger cat state is typically observed.

\section{Entanglement spectrum on cylinders}
\label{sec:ES}

\begin{figure}[h]
\centering
    \subfloat{\includegraphics[width=0.48\columnwidth]{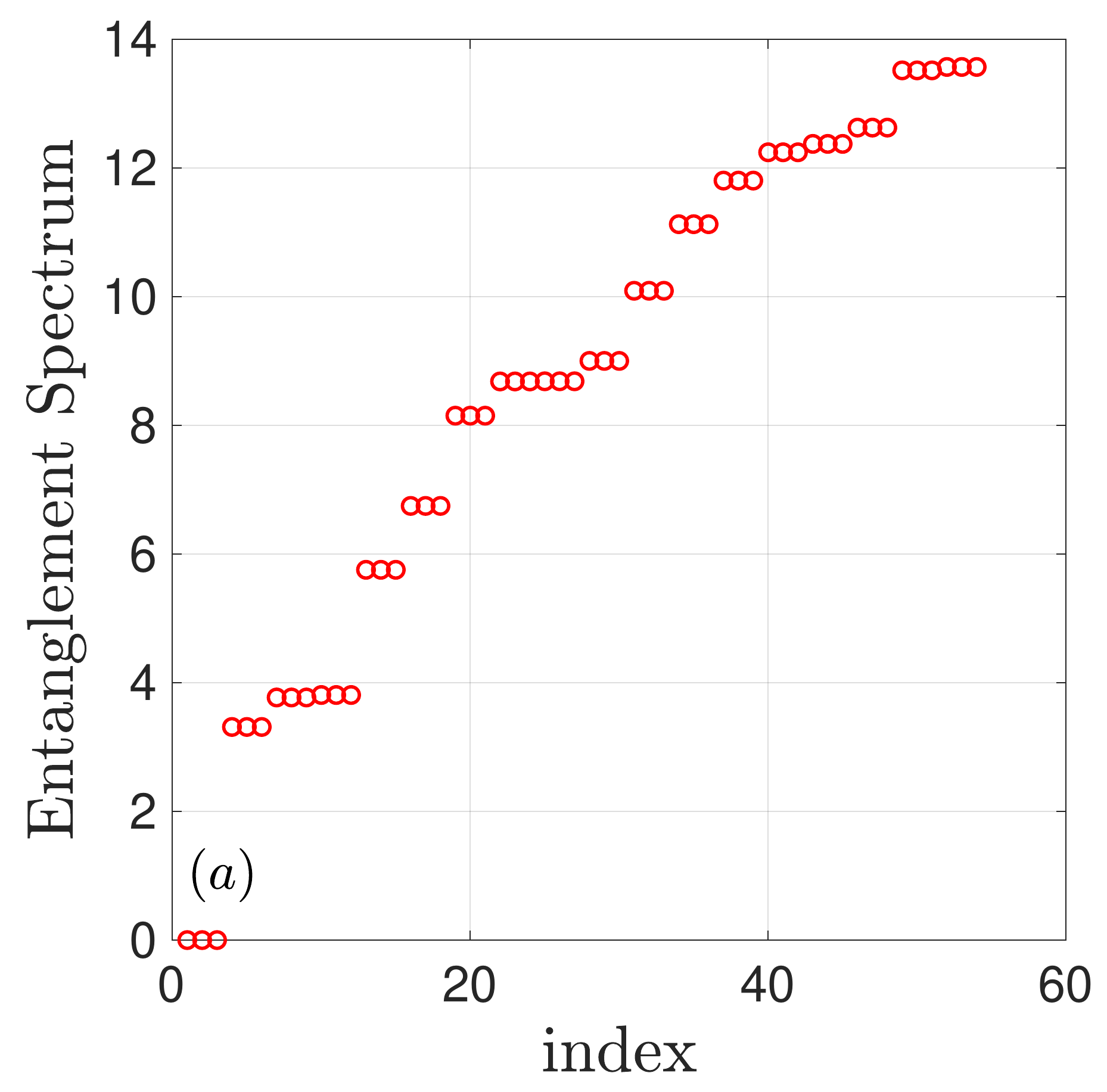}}
    \subfloat{\includegraphics[width=0.475\columnwidth]{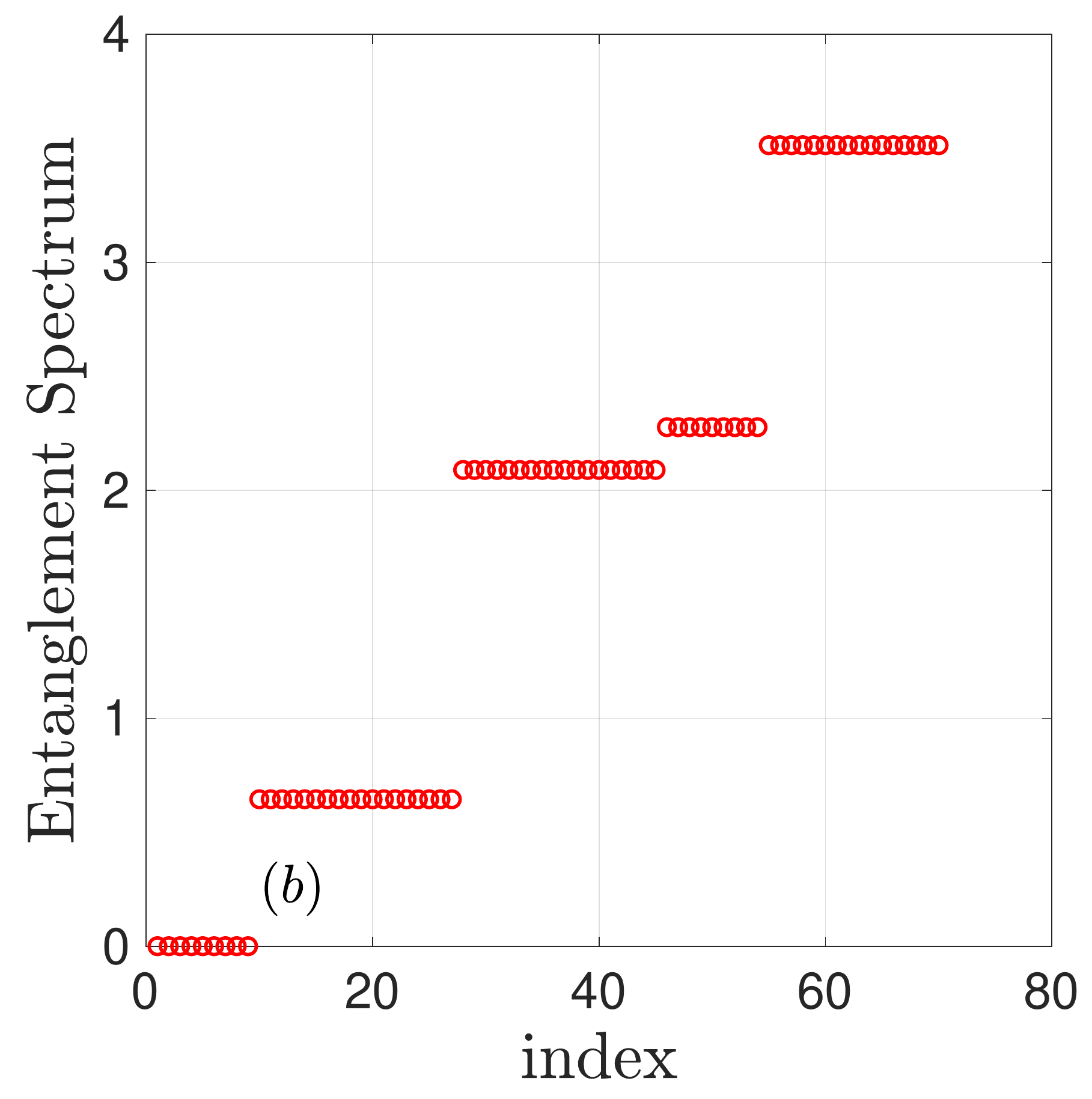}}
\caption{
Entanglement spectrum on finite width cylinders. (a) $L_y=3$, $\chi=800$. (b) $L_y=6$, $\chi=6000$.
}
\label{fig:ES}
\end{figure}

It is known that for chiral topological phases, entanglement spectrum (ES) provides a valuable diagnosis~\cite{Li2008}, which is also accessible from MPS. However, as shown in Fig.~\ref{fig:ES}, the ES for the $\Z_3$ Kitaev model on infinitely long cylinders are heavily degenerate, with no sign of conformal towers. As noted by Ref.~\cite{Shinjo2015} in the context of original Kitaev model, the degeneracy in ES 
can be understood from the conserved quantities $W_p$'s. Since the logic basically follows from Ref.~\cite{Shinjo2015}, we will not elaborate on derivation of the degeneracy in ES but refer to Ref.~\cite{Shinjo2015} for further details.

\section{Degeneracy on the strip geometry}
\label{sec: degeneracy}

\begin{figure}[htb]
  \centering
  \includegraphics[width=0.8\columnwidth]{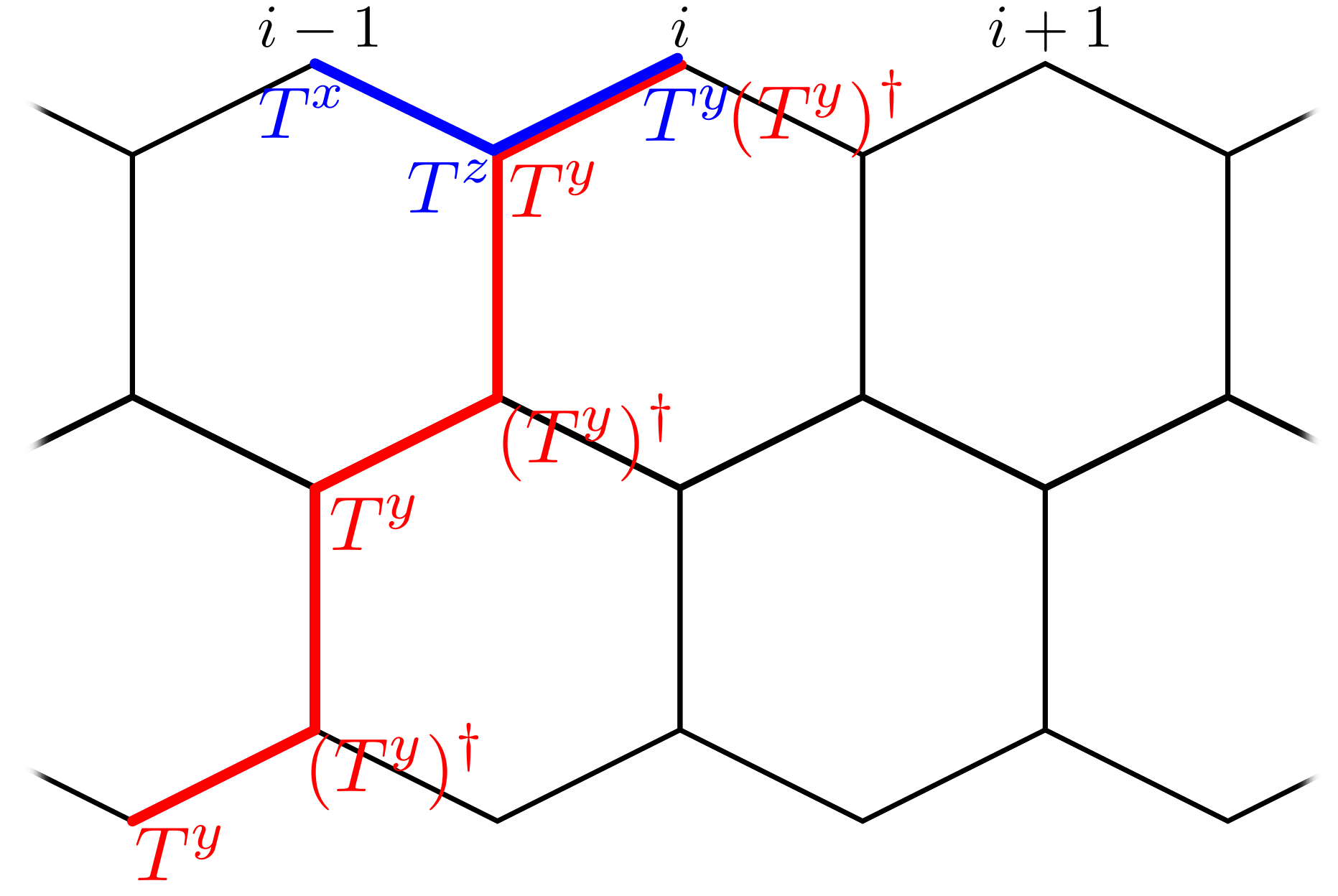}\\
  \caption{The conserved quantities of the $\Z_3$ Kitaev model on a strip geometry are generated by the operators $W_i$ and $\tilde{W}_i$, where $i$ indexes the bivalent vertices at the top of the strip. The operator $W_i$, shown in red, is a product of $T^y$ and $(T^y)^\dagger$ operators, while $\tilde{W}_i$, shown in blue, is a product of $T^x$, $T^y$, and $T^z$. If $L_x$ is even, then each pair of $W_i$ and $\tilde{W}_i$, for even $i$, generates the same operator algebra as a pair of $T^x$ and $T^y$ operators. This implies that each pair enforces a 3-fold degeneracy of the ground state. Note that $\tilde{W}_i$ and $\tilde{W}_{i-1}$ are non-commuting, so to ensure that the pairs are independent, we consider only those at even sites. The subsystem code calculation suggests that it is possible to find $L_x$ independent non-commuting pairs.}
\label{fig: conservedquantities}
\end{figure}

Here, we argue that there is at least a $3^{L_x}$-fold degeneracy of the $\Z_3$ Kitaev model on a strip-like geometry with length $L_x$ in the $\vec{a_1}$-direction and width $L_y$ in the $\vec{a_2}$-direction. To see this, we view the $\Z_3$ Kitaev model as a subsystem code whose gauge group is generated by products of the Hamiltonian terms.
For an introduction to subsystem codes and a more careful description of the code associated to the $\Z_3$ Kitaev model, we refer to Ref.~\cite{Ellison2022}. 

More specifically, we show that the subsystem code has a logical subsystem of dimension $3^{L_x}$. This implies that there are $L_x$ non-commuting pairs of conserved quantities of the $\Z_3$ Kitaev model that satisfy the same commutation relations as $T^x$ and $T^z$. Consequently, the model must have a $3^{L_x}$-fold degeneracy. To support the calculation that follows, we identify $L_x/2$ explicit pairs of non-commuting conserved quantities in Fig.~\ref{fig: conservedquantities}, for a strip of even length $L_x$. This suffices to show that there is at least a $3^{L_x/2}$-fold degeneracy.

We now compute the dimension of the logical subsystem of the subsystem code by comparing the number of physical qutrits (i.e., three-level local degrees of freedom) to the number of stabilized qutrits and gauge qutrits. There are two qutrits for every plaquette. Due to the open boundary conditions, there are an additional $2L_x$ qutrits associated to the edges of the strip. In total, the number of qutrits $\mathbf{N}_Q$ is $\mathbf{N}_Q = 2L_xL_y$.


Next, we count the number of stabilized qutrits and the number of gauge qutrits. The only stabilizers are the conserved quantities $W_p$ for every plaquette, each of which stabilizes a single qutrit. There are $L_x (L_y-1)$ plaquettes, so the number of stabilized qutrits $\mathbf{N}_S$ is $\mathbf{N}_S=L_x (L_y-1)$. There are three gauge generators for every plaquette. There are an additional two for every plaquette on an edge of the strip. Therefore, the number of gauge qutrits $\mathbf{N}_G$ is
\begin{align}
    \mathbf{N}_G = [(3L_x(L_y-1) + 2L_x) - \mathbf{N}_S]/2 = L_xL_y.
\end{align}
Here, we have subtracted $\mathbf{N}_S$, since the stabilizers are generated by the gauge generators, and we have divided by two to account for the fact that the operator algebra of a qutrit has two generators $T^x$ and $T^z$. 

The number of logical qutrits $\mathbf{N}_L$ is then given by the formula 
\begin{align}
    \mathbf{N}_L = \mathbf{N}_Q - \mathbf{N}_S - \mathbf{N}_G = L_x.
\end{align}
Thus, there are $L_x$ logical qutrits, which means that there are $2L_x$ generators of the logical group; each of which is a conserved quantity of the $\Z_3$ Kitaev model. Therefore, we conclude that there is a $3^{L_x}$-fold degeneracy on a strip geometry.

We note that this degeneracy can be lifted by including a single site term in the Hamiltonian for each bivalent vertex at the boundary.


Moreover, for the ground states on strips, under the assumption that the flux on each column is the same, which we verified numerically, we have 
\begin{align}
\Phi_{2,x}\Phi_{2,x+1}\Phi_{2,x+2}&=\Phi_{2,x}\Phi_{2,x+1}^\dag\Phi_{2,x+1}^\dag\Phi_{2,x+2}\\\nonumber
    &=\omega^{nL_y}\cdot\omega^{-nL_y}=1.
\end{align}
where we have used the fact that product of $W_p$ operators in one column equals to $\Phi_{2,x}\Phi_{2,x+1}^\dag$.

\section{Additional data for $L_y=5$ strip}
\label{sec: Ly_5_strip}

\begin{figure}[htb]
    \centering
    \subfloat{\includegraphics[width=0.49\columnwidth]{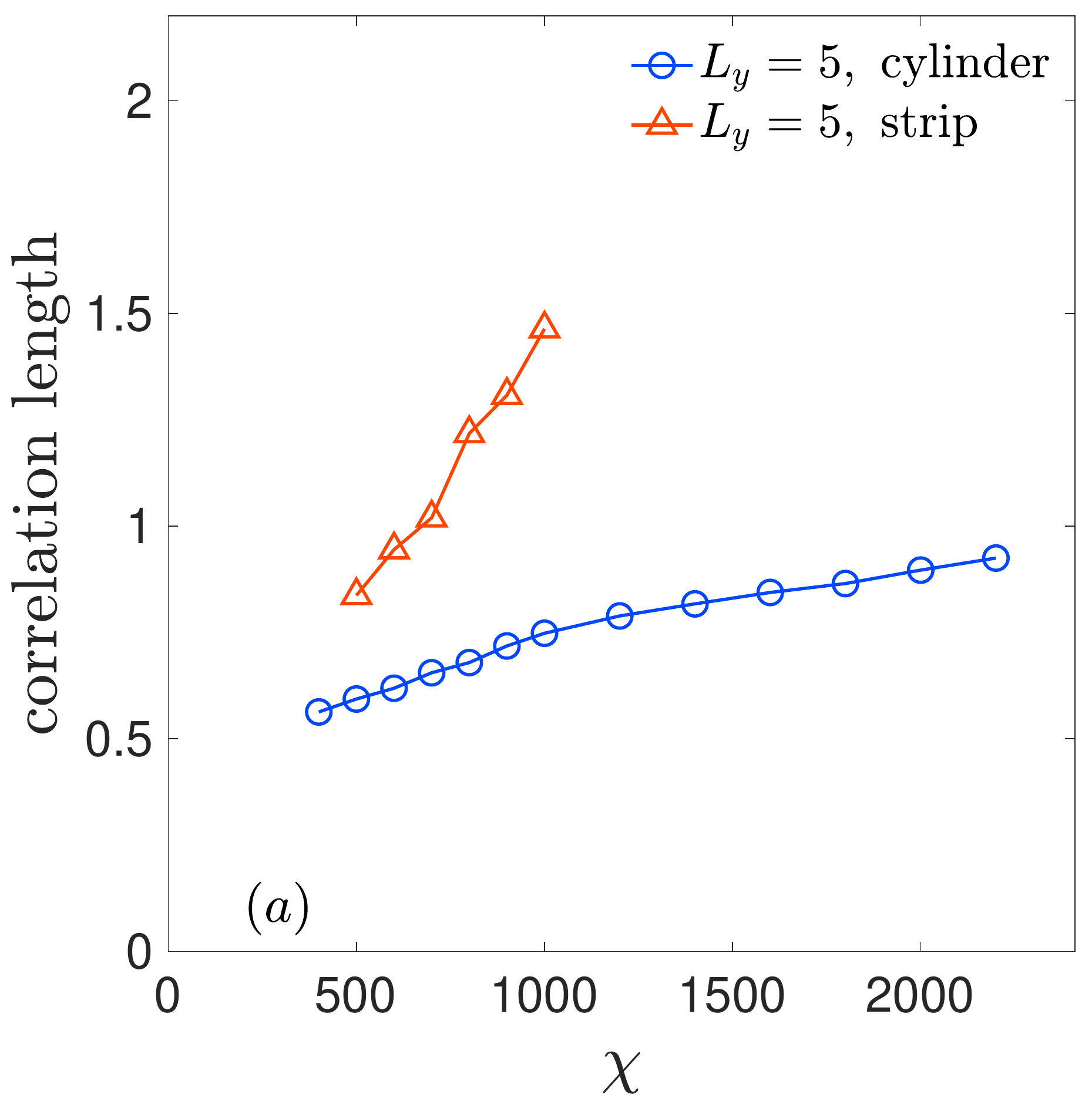}}
    \subfloat{\includegraphics[width=0.49\columnwidth]{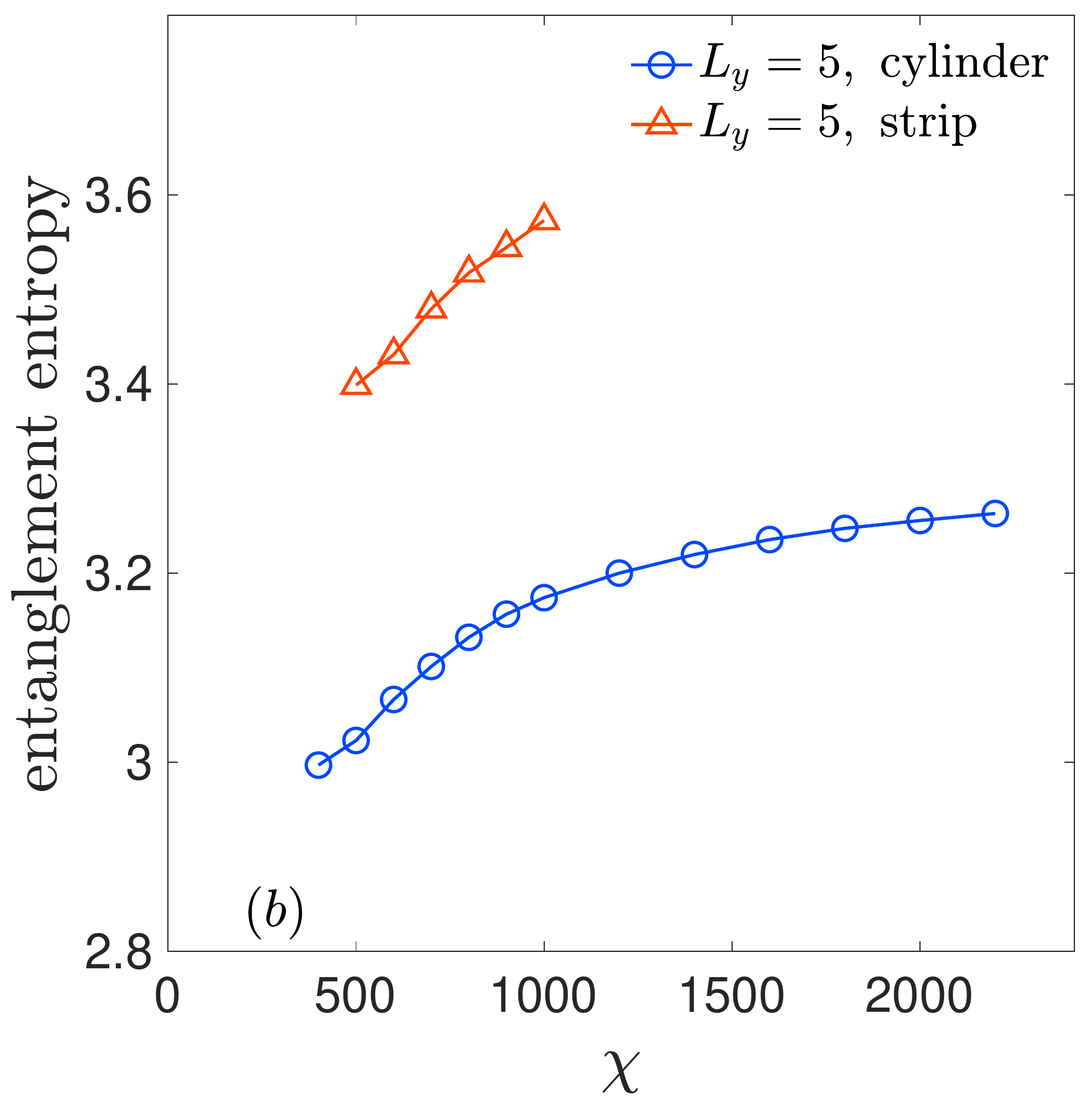}}
    \caption{Comparison of maximal correlation length and entanglement entropy for $L_y=5$ cylinder and strip, shown in (a), (b) respectively. On the strip's boundary, we have $\langle W_p\rangle=\omega$ for all data shown here.}
    \label{fig:Ly_5_strip_cylinder}
\end{figure}

Here we show additional evidence that the critical behavior can also be observed on a $L_y=5$ strip. In Fig.~\ref{fig:Ly_5_strip_cylinder}, we compare the maximal correlation length and entanglement entropy of a $L_y=5$ strip and cylinder. Both correlation length and entanglement entropy grows much quicker (also with larger values) on the strip geometry, without any sign of saturation. This scenario is consistent with the data for $L_y=3,4$ shown in the main manuscript.

\section{Quasi-degeneracy on $L_y=3$ cylinder}
\label{sec: topo_deg}

\begin{figure}[h]
\centering
    \includegraphics[width=0.8\columnwidth]{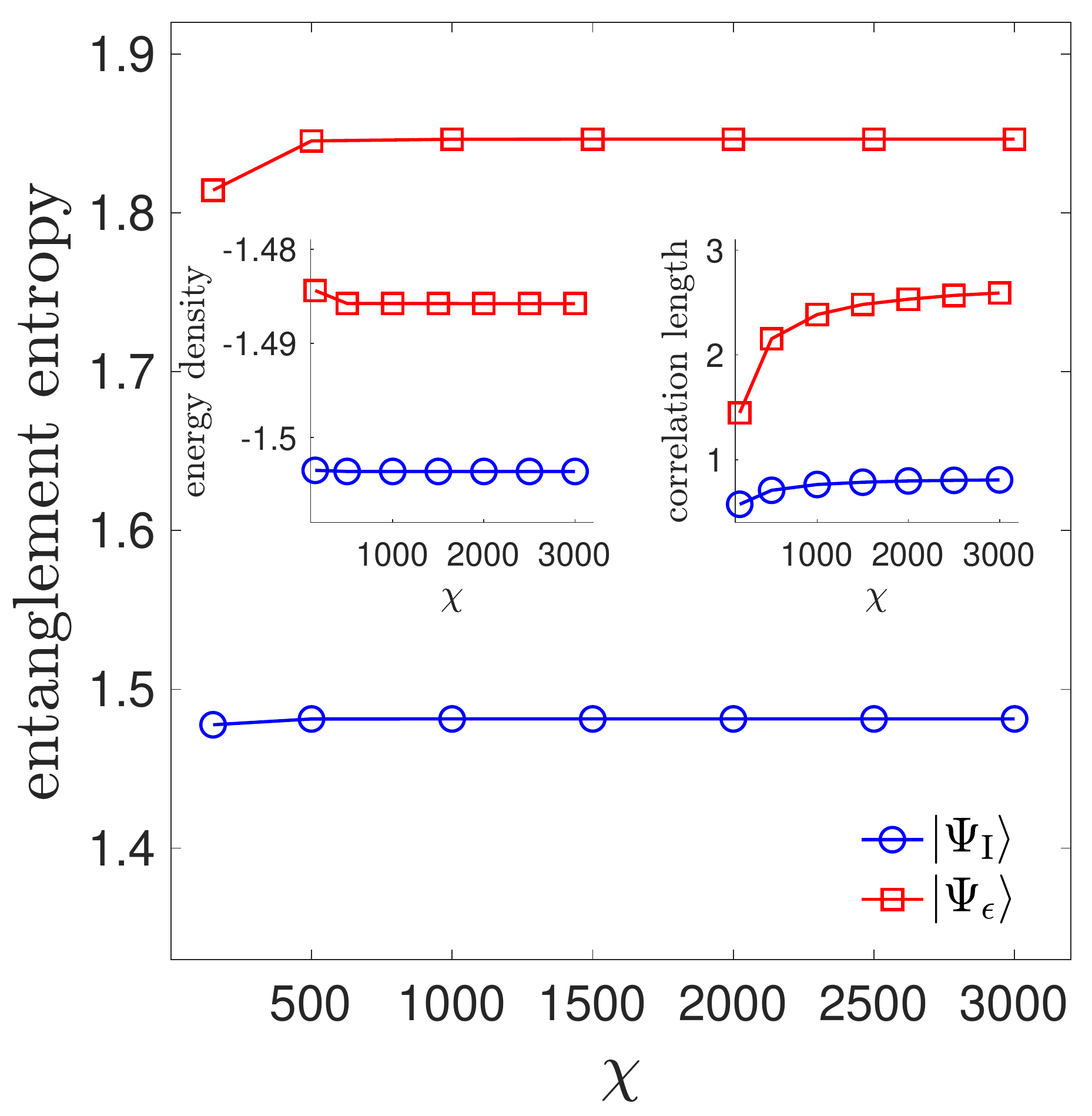}
\caption{Comparison of ground state properties in different topological sectors versus $\chi$ on a YC$3$ cylinder. Here $|\Psi_\mathrm{I}\rangle$ and $\Psi_\epsilon\rangle$ are ground states in distinct topological sectors.
The main panel shows entanglement entropy in the two sectors, with energy density and maximal correlation length shown in the inset.}
\label{fig:TopoDeg}
\end{figure}

Topological order implies (quasi-)degenerate ground states on infinitely long cylinders, which can be found by running iDMRG simulations with suitable initial states~\cite{Cincio2013, Chen2021}. Here, on a YC$3$ cylinder, by randomizing the initial state, we found that the iDMRG simulation has nonzero probability in finding a distinct (excited) state, denoted as $|\Psi_\epsilon\rangle$, while  the actual ground state is $|\Psi_\mathrm{I}\rangle$. $|\Psi_\epsilon\rangle$ is also a simultaneous eigenstate of $W_p$ and $\Phi_2$ with $\langle W_p\rangle=\omega^2$. The overlap per column between  $|\Psi_\mathrm{I}\rangle$ and $|\Psi_\epsilon\rangle$ is $4.2\times 10^{-3}$ with bond dimension $\chi=500$, suggesting that the two states are orthogonal. 

A comparison of $|\Psi_\mathrm{I}\rangle$ and $|\Psi_\epsilon\rangle$ is shown in Fig.~\ref{fig:TopoDeg}. We identify $|\Psi_\epsilon\rangle$ as an approximately degenerate ground state on the finite-circumference cylinder. Because of the finite width, the topological degeneracy would be split by virtual quasiparticle tunneling effects, which explains the energy difference as shown in the inset of Fig.~\ref{fig:TopoDeg}. Moreover, on YC$3$ cylinder, the maximal correlation length of $|\Psi_\epsilon\rangle$ is converging with increasing bond dimension $\chi$ (shown in Fig.~\ref{fig:TopoDeg} inset), implying that the $|\Psi_\epsilon\rangle$ is also gapped, although the value is larger than that of $|\Psi_\mathrm{I}\rangle$. On the other hand, $\ket{\Psi_\epsilon}$ has a higher half cylinder entanglement entropy than $\ket{\Psi_\mathrm{I}}$, with the difference approaching 0.36 with increasing $\chi$. 
 
\bibliography{bibliography}

\end{document}